\documentclass{article}
\usepackage{spconf,amsmath,graphicx}
\usepackage{subfigure}
\usepackage{multirow}
\usepackage{booktabs}
\usepackage{amssymb}
\usepackage{lipsum}

\usepackage{hyperref}
\usepackage{appendix}

\hypersetup{hidelinks,
	colorlinks=true,
	allcolors=black,
	pdfstartview=Fit,
	breaklinks=true}

\title{Emotion Neural Transducer for Fine-Grained Speech Emotion Recognition}
\name{Siyuan Shen$^{1,2,3\dag}$ \qquad Yu Gao$^{3\ast}$ \qquad Feng Liu$^{1,2}$ \qquad Hanyang Wang$^{1,2}$ \qquad Aimin Zhou$^{1,2\ast}$ \thanks{This paper is funded by the Science and Technology Commission of Shanghai Municipality (Grant No. 22511105901) and the Beijing Key Laboratory of Behavior and Mental Health, Peking University. \\
$^{\dag}$ Work done while intern at Midea Group. \\
$^{\ast}$ Corresponding author. gaoyu11@midea.com, amzhou@cs.ecnu.edu.cn}}

\address{$^{1}$Shanghai Institute of AI for Education, East China Normal University, Shanghai, China\\
	     $^{2}$School of Computer Science and Technology, East China Normal University, Shanghai, China\\
      $^{3}$AI Innovation Center, Midea Group, Shanghai, China}
\begin{document}
%\ninept
%
\maketitle
\begin{abstract}
    The mainstream paradigm of speech emotion recognition (SER) is identifying the single emotion label of the entire utterance. This line of works neglect the emotion dynamics at fine temporal granularity and mostly fail to leverage linguistic information of speech signal explicitly. In this paper, we propose Emotion Neural Transducer for fine-grained speech emotion recognition with automatic speech recognition (ASR) joint training. We first extend typical neural transducer with emotion joint network to construct emotion lattice for fine-grained SER. Then we propose lattice max pooling on the alignment lattice to facilitate distinguishing emotional and non-emotional frames. To adapt fine-grained SER to transducer inference manner, we further make blank, the special symbol of ASR, serve as underlying emotion indicator as well, yielding Factorized Emotion Neural Transducer. For typical utterance-level SER, our ENT models outperform state-of-the-art methods on IEMOCAP in low word error rate. Experiments on IEMOCAP and the latest speech emotion diarization dataset ZED also demonstrate the superiority of fine-grained emotion modeling. Our code is available at \href{https://github.com/ECNU-Cross-Innovation-Lab/ENT/}{https://github.com/ECNU-Cross-Innovation-Lab/ENT}.
\end{abstract}
\begin{keywords}
Speech emotion recognition, speech emotion diarization, automatic speech recognition
\end{keywords}
\section{Introduction}

Speech emotion recognition (SER) aims to identify emotional states of human speech signals. Many works follow the recipe of classifying the whole utterance into single emotion category \cite{pepino2021emotion, liu2023speech, kakouros2023speech, shen2023mingling, li23m_interspeech, sun23d_interspeech}. Fueled by various datasets containing emotional labels at utterance level \cite{busso2008iemocap, poria2019meld}, such sequence-to-label methods have made great progress in recent years. However, emotional states inherently exhibit diverse temporal dynamics, often leading to alternating shifts between emotional and non-emotional states within a single utterance \cite{han2018towards}. Consequently, recognizing emotions at a fine temporal granularity is desirable for better emotion understanding \cite{wang2023speech, li2022dilated}.

For fine-grained SER, another line of previous works consider this task as a sequence-to-sequence problem. These methods can be thought of as aligning frames and emotional labels in a weakly supervised manner. Frame-wise methods simply assign overall emotional label to each frame \cite{mirsamadi2017automatic} while segment-wise methods identify emotional regions according to contribution of salient parts with attention mechanism \cite{mao2020advancing}. Connectionist temporal classification (CTC) methods \cite{graves2006connectionist} first construct emotion sequence heuristically and then align emotionally relevant segments within the utterance automatically \cite{han2018towards}. Though driven by the common motivation for fine-grained SER, these approaches only consider acoustic information of speech signals and evaluate performance at utterance level. Thanks to the latest proposed benchmark of speech emotion diarization \cite{wang2023speech}, the frontier of distinguishing emotions at a fine temporal granularity is to be uncovered.

Motivated by the nature of neural transducer for sequence alignment conditioning on both linguistic tokens and acoustic units \cite{graves2012sequence,zhang2020transformer}, we explore SER and fine-grained SER based on transducer models with automatic speech recognition (ASR) joint training. To date, recent paradigms for joint SER and ASR at utterance level include cascading off-the-shelf ASR model \cite{Chen2022CTARNNCA} as well as adopting multi-task learning framework, where intermediate \cite{feng2020end,li2022fusing} or task-specific output layers \cite{cai2021speech} are supervised by CTC loss. Despite the huge success of transducer family in the field of ASR \cite{graves2012sequence,zhang2020transformer,chen2022factorized}, existing extension on RNN-T for additional SER functionality solely focuses on modifying target transcriptions with emotion tags \cite{kons2022extending}. Moreover, these ASR-based methods view SER as a typical utterance-level classification problem, disregarding the temporal granularity of emotion.

In this paper, we aim to bridge the gap between ASR-based SER and fine-grained SER, allowing generating rich transcripts along with emotion synchronously. We propose Emotion Neural Transducer (dubbed ENT) for fine-grained speech emotion recognition with ASR joint training. We first build the emotion joint network upon the typical acoustic encoder and vocabulary predictor and thus enable modeling emotion categorical distribution through the alignment lattice as standard neural transducer \cite{graves2012sequence}. Since fine-grained SER operates under a a weakly-supervised learning paradigm, we propose lattice max-pooling loss for the emotion lattice to distinguish emotional and non-emotional timestamps automatically. Motivated by the inference manner of neural transducer, we further extend emotion neural transducer to the factorized variant (called FENT). The key concept behind FENT is to utilize the blank symbol as both a time separator and an underlying indicator of emotion. Specifically, we disentangle emotion and blank prediction from vocabulary prediction with separate predictors and share the predictor for both blank and emotion prediction. Our proposed ENT models outperform previous state-of-the-arts on the benchmark IEMOCAP dataset with low word error rate. Moreover, we validate fine-grained emotion modeling with ASR on the newly proposed emotion diarization dataset ZED.

% The core idea is to make the blank serve as both time separator and underlying emotion indicator. Inspired by recent advances in language adaptation \cite{chen2022factorized}, we first disentangle emotion and blank prediction from vocabulary prediction. Specifically, the vocabulary predictor is dedicated to predict label vocabulary excluding blank while the blank predictor produces blank probability. To integrate both acoustic and linguistic tokens, the emotion joint network is then built upon the acoustic encoder and blank predictor, shared by the blank joint network. In this way, the joint network models the underlying emotion categorical distribution through the alignment lattice as standard neural transducer \cite{graves2012sequence}. Since fine-grained speech emotion recognition follows weakly-supervised learning paradigm, we propose a simple lattice max-pooling loss for the emotion lattice to distinguish emotional and non-emotional timestamps automatically. Our proposed FENT outperforms previous state-of-the-arts on the benchmark IEMOCAP dataset with low word error rate. Moreover, we pioneer the exploration of fine-grained SER with ASR joint training and validate on the newly proposed emotion diarization benchmark. %This paper is funded by  Special Project on Support for Regulative Intelligence Technology under the Shanghai Science and Technology Innovation Plan 2022 on "Research on the Theory, Assessment System and Prototype System for Enhancing Emotional Cognition in the General Population" (Grant No. 22511105901)

% \section{Methodology}

\section{Neural Transducer}
Standard neural transducer consists of three components, the acoustic encoder, prediction network and joint network \cite{graves2012sequence, zhang2020transformer}. Considering the acoustic input $x$ with duration $T$ and target label sequence $y$ with length $U$, the acoustic encoder takes acoustic features $x_{\le t}$ as input and produces hidden features $h_t$ for each timestamp. The prediction network generates label representations $g_u$ conditioning on previous tokens $y_{\le u}$. The joint network integrates the outputs of acoustic encoder and prediction network as $z_{t,u}$ to compute vocabulary label distribution. The procedure can be formulated as
\begin{equation}
\centering
\begin{aligned}
    h_t &= \mathrm{Encoder}(x_{\le t})\\
    g_u &= \mathrm{Predictor}(y_{\le u})\\
    z_{t,u} &= \mathrm{Joint}(h_t,g_u)\\
\end{aligned}
\label{eq:transducer}
\end{equation}

Then the probability of next token can be computed as
\begin{equation}
    P(y_{u+1}\mid x_{\le t},y_{\le u }) = \mathrm{softmax}(z_{t,u}).
\end{equation}

To address the length difference between acoustic features $x$ and token sequences $y$, transducer models add a special blank symbol to the vocabulary for alignment and optimize log probability over all possible alignments as

\begin{equation}
    \mathcal{L}_{trans}=-\mathrm{log}\sum_{\alpha \in \beta^{-1}(y)} (P(\alpha \mid x)),
\end{equation}

where $\alpha$ denotes the alignment, each containing $T+U$ tokens and $\beta$ is the mapping from alignment to target sequence by removing blank symbols.

\section{Emotion Neural Transducer}
In this section, we present two key components of ENT and subsequently extend it to its factorized variant FENT. First, we construct the emotion joint network to integrate representations from the encoder and predictor to yield emotion lattice. To further enhance emotional and non-emotional awareness at temporal granularity, we then devise lattice max pooling loss to the generated emotion lattice. Next, we make the blank symbol work as an emotion indicator for FENT by disentangling blank from vocabulary and meanwhile sharing the same predictor for both blank and emotion prediction.

\subsection{Joint Emotion Prediction}
% Our intuition for fine-grained SER with neural transducer is based on its inference manner. At each timestamp, the standard transducer model consumes one frame and then outputs multiple non-blank tokens until the blank is emitted. We assume blank symbol as accumulation of acoustic and linguistic information and thus we allocate temporal emotion awareness to blank representations. As shown in Figure, we let emotion and blank prediction share the same predictor and use emotion-specific joint network $\mathrm{joint^E}$ for emotion prediction. Considering the complex coordinating roles of transducer components \cite{ghodsi2020rnn, meng2021internal}, we take initiatives to preserve the modularity of transducer for ASR capability. Specifically, we add hidden features as hidden fusion rather than previous log probability domain methods \cite{chen2022factorized}.

\begin{figure}[!t]
    \centering
    \includegraphics[width=0.8\linewidth]{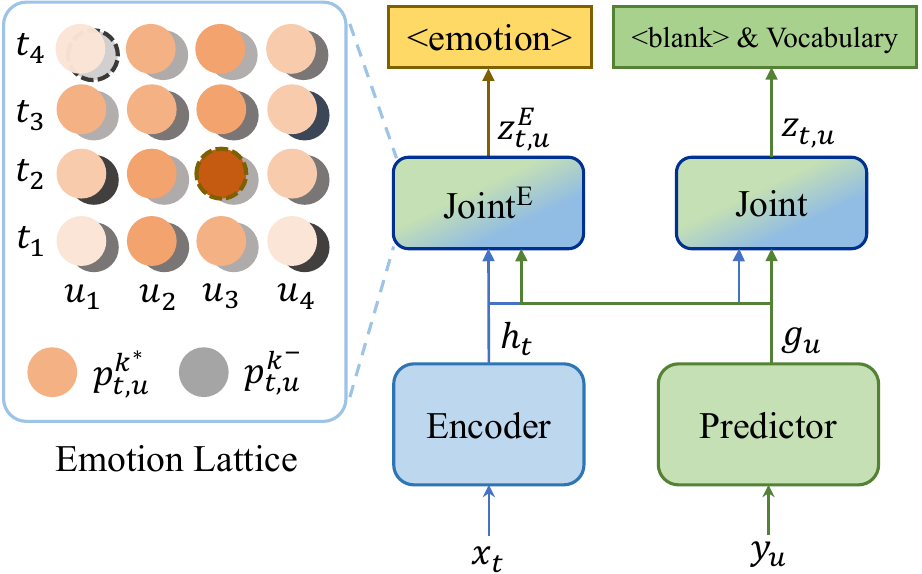}
    \caption{Emotion Neural Transducer.}
    \label{fig1}
\end{figure}

To integrate both acoustic and linguistic tokens for emotion recognition at a fine temporal granularity, we build emotion joint network upon the typical acoustic encoder and vocabulary predictor (Figure \ref{fig1}). Formally, the emotion representation $z^E_{t,u}$ given speech and text history can be obtained by substituting $\mathrm{joint^E}$ into Equation \ref{eq:transducer}. Similar to standard neural transducer modeling sequence alignment via lattice \cite{graves2012sequence}, our emotion joint network models emotion emission probability through the $T \times U$ alignment lattice. As shown in Figure \ref{fig1}, each node $p_{t,u}$ denotes the emotion probability distribution having output $u$ tokens by frame $t$, where $p_{t,u}^{k}$ is the probability of emotion $k$ with darker color indicating higher probability, orange/grey denoting emotional/non-emotional.

\subsection{Lattice Max Pooling}
 Given the utterance-level emotion label $k^*$, our goal is to identify the emotional and non-emotional frames automatically through the lattice. Inspired by max pooling loss used in keyword spotting \cite{hou2020mining,wang2023wekws}, we extend the frame-level max pooling loss on the emotion lattice, thus leveraging acoustic and linguistic alignment. For each utterance, we select the node with the highest predicted posterior probability of target emotion $p_{t,u}^{k^*}$ and the node with the minimum non-emotional or neutral category probability $p_{t,u}^{k^-}$. In Figure \ref{fig1}, the selected nodes are indicated by dashed borderline. Our proposed lattice max pooling loss can be formulated as 

\begin{equation}
    \mathcal{L}_{lattice}=-\mathrm{log} \max_{t,u} (p_{t,u}^{k^*})-\mathrm{log}\min_{t,u} (p_{t,u}^{k^-}).
\end{equation}

From the view of positive and negative samples as max pooling loss, the first term optimizes the most positive frame while the second term selects the hardest negative sample through the emotion lattice. See Appendix \ref{app:variants} for variants.

To maintain the capability of conventional SER at utterance level, we do mean pooling for the representations from acoustic encoder and predictor, optimized by cross entropy loss $\mathcal{L}_{emotion}$ at utterance level.

% Then the overall objective of ENT can be rewritten as

% \begin{equation}
% \mathcal{L}_{ENT}=\mathcal{L}_{trans}+\mathcal{L}_{lattice}+\mathcal{L}_{emotion}.
% \end{equation}

\begin{figure}[!t]
    \centering
    \includegraphics[width=0.8\linewidth]{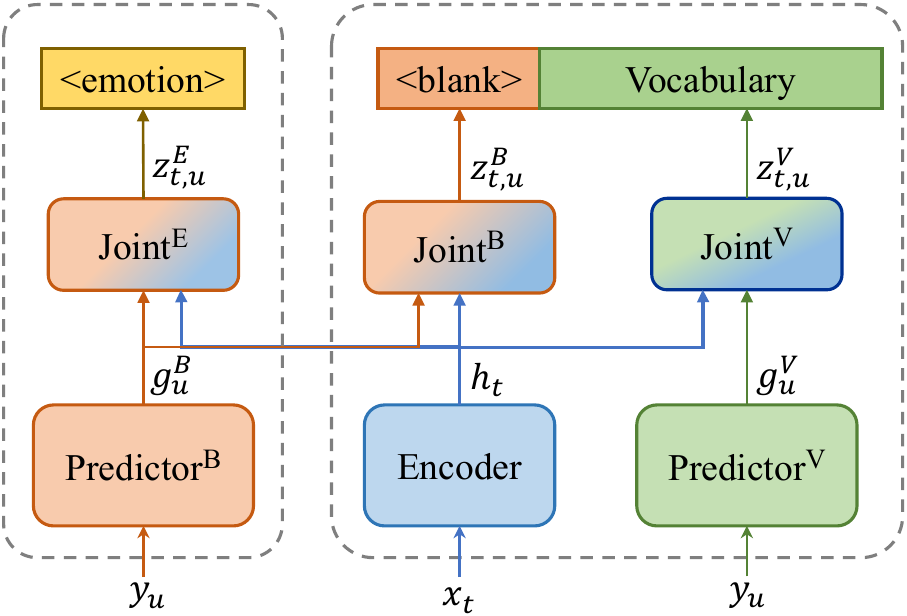}
    \caption{Factorized Emotion Neural Transducer.}
    \label{fig2}
\end{figure}

\subsection{Factorized Emotion Neural Transducer}
Our intuition for the factorized variant is based on the natural inference manner of neural transducer. At each timestamp, the standard transducer model consumes one frame and then outputs multiple non-blank tokens until the blank is emitted. We assume the blank symbol as accumulation of both acoustic and linguistic information and thus we allocate temporal emotion awareness to blank representations. Inspired by recent advances in language model adaptation \cite{chen2022factorized, variani2020hybrid}, we first disentangle blank from vocabulary prediction by using two separate predictors. The overall architecture of FENT is described in Figure \ref{fig2}. Specifically, the vocabulary $\mathrm{predictor^V}$ is dedicated to predicting label vocabulary representations $g^V_u$ excluding blank while the blank predictor $\mathrm{predictor^B}$ produces blank representations $g^B_u$ as the right part of Figure \ref{fig2}. Then the corresponding joint network fuses acoustic features $h_t$ with predictor outputs similar to Equation \ref{eq:transducer}, yielding $z^V_{t,u}$ and $z^B_{t,u}$ respectively. The whole vocabulary label distribution can be computed by softmax and concatenation as
\begin{equation}
    P(y_{u+1}\mid x_{\le t},y_{\le u }) = \mathrm{softmax}([z^B_{t,u}; z^V_{t,u}]).
\end{equation}

To bias the blank symbol towards emotion, we employ a shared predictor for emotion and blank prediction and adopt aforementioned emotion joint network as shown in left part of Figure \ref{fig2}. For each time step during inference, the acoustic encoder takes one frame as input and the vocabulary predictor outputs the most probable tokens iteratively until the blank is emitted. At this point, the emotion joint network fuses acoustic and blank representation to predict current emotion.

\section{Experiments}
In this section, we first demonstrate the superiority of ENT models on the benchmark dataset IEMOCAP for utterance-level SER. Next we validate the capability of fine-grained speech emotion recognition on the speech emotion diarization dataset ZED and meanwhile ablate key components.

\subsection{Experimental Setup}
\textbf{Dataset and evaluation.} Interactive emotional dyadic motion capture database (IEMOCAP) \cite{busso2008iemocap} is a widely-used benchmark SER dataset, where each utterance is annotated with the transcript and single emotion category label. We adopt leave-one-session-out 5-fold cross-validation, following the typical evaluation protocol. The unweighted accuracy (UA) and weighted accuracy (WA) for utterance-level SER are computed by averaging the results obtained from the 5 folds. The average word error rate (WER) across the 5 folds is reported to measure ASR performance.

Zaion Emotion Dataset (ZED) \cite{wang2023speech} is a recently proposed dataset for fine-grained SER, named as speech emotion diarization, including 180 utterances annotated with emotional boundaries for each. It is worth noting that due to its limited scale, ZED is primarily suitable for evaluating the fine-grained SER capability rather than serving as a comprehensive training set in a fully supervised manner.  Thus we train our ENT models on IEMOCAP and validate on ZED. We adopt emotion diarization error rate (EDER) for fine-grained SER, which assesses the temporal alignment between predicted emotion intervals and the actual emotion intervals. Lower EDER indicates better fine-grained SER ability.

\textbf{Implementation Details.} We take wav2vec 2.0 Base \cite{baevski2020wav2vec} as feature extractor for input speech signals, where the pre-trained model is frozen for training efficiency and the features from different layers are performed weighted sum in line with SUPERB \cite{yang2021superb}. The acoustic encoder and the predictors are one-layer LSTM with a hidden dimension of 640. The joint network combines features of encoder and predictor by addition operation, followed by a linear layer.
% We adopt AdamW optimizer with learning rate of $5\times10^{-4}$ and decay rate of 0.1. We train at batch size of 2 for 50 epochs with 5 epochs of linear warm-up.

\begin{table}[!t]
\centering
\small
\begin{tabular}{@{}ll|cc@{}}
\toprule
\textbf{Method} & \textbf{Year} & \textbf{WA (\%)} & \textbf{UA (\%)} \\ \midrule
Wav2vec2-PT \cite{pepino2021emotion}     & 2021          & 67.90            & -                \\
Corr Attentive \cite{liu2023speech}  & 2023          & -                & 70.01            \\
DCW+TsPA \cite{kakouros2023speech}     & 2023          & 72.08            & 72.17            \\
Shiftformer \cite{shen2023mingling}     & 2023          & 72.10            & 72.70            \\
MSTR \cite{li23m_interspeech}          & 2023          & 70.60            & 71.60            \\
EmotionNAS \cite{sun23d_interspeech}      & 2023          & 69.10            & 72.10            \\ \midrule
ENT (ours)      & 2023          & \textbf{72.43}   & \textbf{73.88}   \\
FENT (ours)     & 2023          & 71.84            & 72.37            \\ \bottomrule
\end{tabular}
\caption{Comparison with utterance-level SER methods using wav2vec 2.0 as feature extractor on IEMOCAP.}
\label{table:SER}
\end{table}

\subsection{Utterance-level Speech Emotion Recognition}
\textbf{Comparison with state-of-the-arts.} We compare our proposed models with recent state-of-the-art methods in Table \ref{table:SER}. ENT outperforms all the strong baselines, showing the effectiveness of leveraging linguistic information and fine-grained temporal modeling. While FENT achieves competitive results as well, the factorization technique degrades its utterance-level SER performance slightly compared with ENT just as its counterpart in language adaptation \cite{chen2022factorized}. This phenomenon indicates that factorization of predictor partially compromises the integrity of whole vocabulary modeling, resulting in inferior representation for utterance-level discrimination.

\textbf{Comparison with ASR-based methods.} We evaluate the SER and ASR performance in Table \ref{table:ASR}. For fair comparison, all the methods take features from self-supervised or ASR pre-trained models. Although ASR joint training enables the model to predict emotions along with transcriptions, previous attempts fail to balance the mutual influence between ASR and SER. Taking RNN-T method as an example, appending a special emotion tag to the target text is conductive to the original ASR output manner, yet deteriorating SER ability (only 58.2\% WA). In contrast, the family of ENT attains better performance in both ASR and SER (+3\% UA and meanwhile -0.7\% WER). Notably, the top performance of FENT in WER validates the effectiveness of factorization of emotion from vocabulary, preserving the modularity of the transducer for ASR capability \cite{ghodsi2020rnn, meng2021internal} while endowing SER capability.

\begin{table}[!h]
\centering
\small
\begin{tabular}{@{}l|l|cc@{}}
\toprule
\textbf{Type}        & \textbf{Method}          & \textbf{WA (\%)} & \textbf{WER (\%)} \\ \midrule
\multirow{3}{*}{CTC} & e2e-ASR \cite{feng2020end}                 & 68.60            & 35.70             \\
                     & wav2vec 2.0+co-attention \cite{li2022fusing} & 63.40            & 32.70             \\ \midrule
RNN-T                & Emotion tag \cite{kons2022extending}             & 58.20            & 26.70             \\ \midrule
\multirow{2}{*}{ENT} & ENT (ours)               & \textbf{72.43}   & 26.47             \\
                     & FENT (ours)              & 71.84            & \textbf{25.99}    \\ \bottomrule
\end{tabular}
\caption{Comparison with ASR-based utterance-level SER methods on IEMOCAP.}
\label{table:ASR}
\end{table}

\subsection{Fine-Grained Speech Emotion Recognition}
Table \ref{table:SED} is split into 3 parts to compare with frame-wise methods and ENT variants. It is noteworthy that the weak supervision paradigm based on utterance-level annotation and absence of an appropriate training set makes fine-grained SER validation on SED benchmark extremely challenging. Interestingly, standard ENT without lattice loss, though utilizing text information explicitly, lags behind frame-wise baseline by nearly 3\% EDER, suffering from degraded ASR capability as well as imperfect transcripts. Thanks to disentangling emotion and blank from vocabulary prediction, our FENT reaches much lower EDER (about -4.6\%) while enjoying speech transcription functionality along with fine-grained emotion. 

\begin{table}[]
\centering
\small
\begin{tabular}{@{}l|cc|cc@{}}
\toprule
\multirow{2}{*}{\textbf{Method}}        & \multicolumn{2}{c|}{\textbf{IEMOCAP}}              & \multicolumn{2}{c}{\textbf{ZED}}                       \\ \cmidrule(l){2-5} 
                                        & \textbf{UA $\uparrow$} & \textbf{WER $\downarrow$} & \textbf{EDER $\downarrow$} & \textbf{WER $\downarrow$} \\ \midrule
Frame-wise                              & 68.43                  & -                         & 59.73                           & -                         \\ \midrule
ENT                                     & \textbf{73.88}         & 26.47            & 56.60                      & 39.37                     \\
-w/o $\mathcal{L}_{lattice}$            & 71.76                       & 26.06                          & 62.47                      & 39.19                     \\
\quad -w. $\mathcal{L}^{T}_{lattice}$   & 73.11                       & 26.42                          & 61.88                           & 39.39                          \\
\quad -w. $\mathcal{L}^{U}_{lattice}$   & 69.86                       & 26.14                          & 61.40                           & \textbf{38.82}                          \\
\quad -w. $\mathcal{L}^{all}_{lattice}$ & 71.85                  & 26.19                     & 61.12                      & 39.28                     \\
\quad\quad -w. mixing & -                  & -                     & 52.76*                      & 42.54*                      \\
-w. BPE                        & 71.37                       & 30.13                          & 67.68                           & 47.42                          \\ \midrule
FENT                                    & 72.37                  & \textbf{25.99}                     & \textbf{55.07}             & 39.34            \\
-w/o $\mathcal{L}_{lattice}$            & 71.84                  & 26.69                     & 60.86                      & 39.14                     \\
\quad -w. $\mathcal{L}^{T}_{lattice}$   & 72.52                       & 26.28                          & 59.18                           & 39.48                          \\
\quad -w. $\mathcal{L}^{U}_{lattice}$   & 69.67                       & 26.23                          & 60.86                           & 39.17                         \\
\quad -w. $\mathcal{L}^{all}_{lattice}$ & 69.61                  & 26.18                     & 59.38                      & 39.11                      \\
\quad\quad -w. mixing & -                  & -                     & 54.41*                      & 39.93*                      \\
-w. BPE                        & 70.33                       & 30.96                          & 65.63                           & 47.26                          \\ \bottomrule
\end{tabular}
\caption{Comparison of ENT varaints performance on IEMOCAP and ZED. * denotes training models with concatenated IEMOCAP audio segments like \cite{wang2023speech} and $\mathcal{L}^{all}_{lattice}$.}
\label{table:SED}
\end{table}

\subsection{Ablation Studies}
We investigate key components of ENT models in Table \ref{table:SED}. Overall, FENT architecture excels at fine-grained SER on SED regardless of $\mathcal{L}_{lattice}$ while ENT obtains better UA in typical utterance-level SER. Compared with character units, text encoded with byte-pair encoding (BPE) degrades WER as well as emotion recognition performance significantly, which may be attributed to vocabulary sparsity for relatively small SER dataset, further yielding negative mutual impact of speech and emotion recognition. We then compare our lattice max pooling to some straightforward variants, where $\mathcal{L}^{T}_{lattice}$ selects the entire timestamp (target row of emotion lattice in Figure \ref{fig1}) while $\mathcal{L}^{U}_{lattice}$ selects the target token column. And $\mathcal{L}^{all}_{lattice}$ applies supervision on the whole emotion lattice. We can observe that $\mathcal{L}^{T}_{lattice}$ achieves on-par performance as original $\mathcal{L}_{lattice}$, signifying the importance of temporal localization. More importantly, the improvement of models with lattice max pooling on IEMOCAP also verifies that fine-grained emotion modeling helps utterance-level SER. Moreover, improvement by mixing different audio segments shows compatibility of our lattice loss to supervised data.

\section{Conclusion}
In this paper, we present Emotion Neural Transducer models for fine-grained speech emotion recognition, with a favorable capability of predicting transcripts along with emotion at fine temporal granularity for practice. We hope our work will draw more attention from the community toward more comprehensive fine-grained emotion benchmarks.

\vfill\pagebreak

\footnotesize
\bibliographystyle{IEEEbib}
\bibliography{main}

\clearpage
\appendix
\section{Variants of Lattice Max Pooling}
\small
\label{app:variants}
As mentioned in our experiment, the lattice max pooling loss can be extended to some variants based on the groups of selected node and the supervision manner. We define the indices of the nodes with the highest predicted probability of the target emotion as $t^*$ and $u^*$, and the indices of the nodes with the minimum non-emotional probability as $t^-$ and $u^-$.
\begin{equation}
\begin{aligned}
    &t^*,u^* = \arg\max_{t,u} (p_{t,u}^{k^*}),\\
    &t^-,u^- = \arg\min_{t,u} (p_{t,u}^{k^-}).\\
\end{aligned}
\end{equation}

\begin{figure}[!ht]
    \centering
    \subfigure[Temporal Lattice Max Pooling $\mathcal{L}^{T}_{lattice}$.]{
          \includegraphics[width=0.35\linewidth]{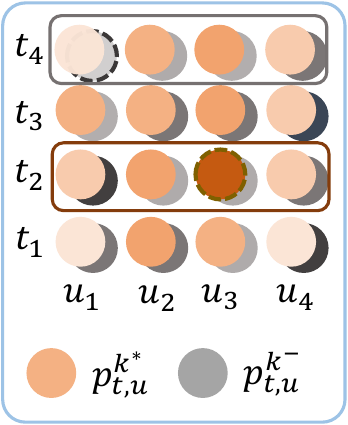}
            %\caption{Original.}
            \label{fig:lattice_T}
    }
    \hfil
    \subfigure[Token Lattice Max Pooling $\mathcal{L}^{U}_{lattice}$.]{
          \includegraphics[width=0.35\linewidth]{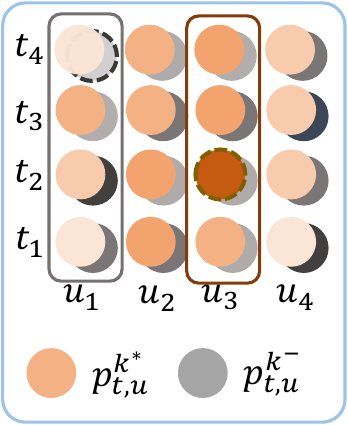}
            %\caption{Unidirection.}
            \label{fig:lattice_U}
    }
    \caption{Temporal and Token Lattice Max Pooling Loss.}
    \label{fig:lattice}
\end{figure}

\textbf{Temporal Lattice Max Pooling $\mathcal{L}^{T}_{lattice}$} (see Figure \ref{fig:lattice_T}) first selects the nodes within the entire timestamp row instead of a single node and then calculates the loss as follows
\begin{equation}
    \mathcal{L}_{lattice}^T=- \sum_{u} \mathrm{log}(p_{t^*,u}^{k^*})-\sum_{u} \mathrm{log}(p_{t^-,u}^{k^-}).
\end{equation}

\textbf{Token Lattice Max Pooling $\mathcal{L}^{U}_{lattice}$} (see Figure \ref{fig:lattice_U}) first selects the nodes within the entire token column and then calculates the loss as follows
\begin{equation}
    \mathcal{L}_{lattice}^U=- \sum_{t} \mathrm{log}(p_{t,u^*}^{k^*})-\sum_{t} \mathrm{log}(p_{t,u^-}^{k^-}).
\end{equation}

\begin{figure}[!ht]
    \centering
    \includegraphics[width=0.45\linewidth]{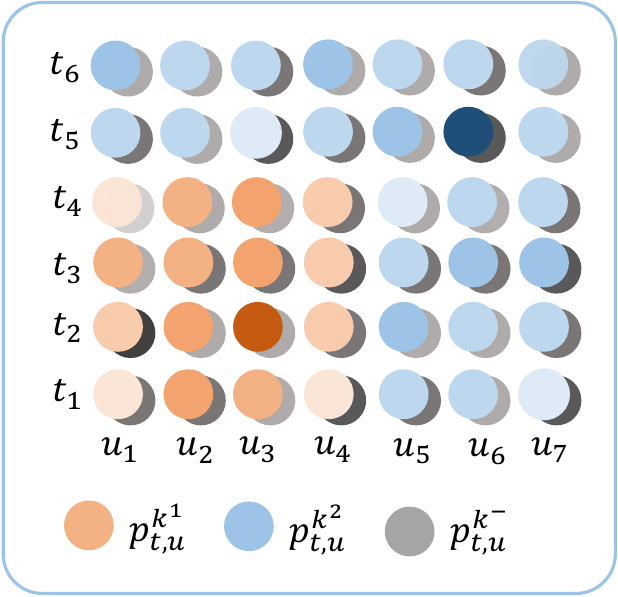}
    \caption{Mixing on Emotion lattice.}
    \label{fig:mixing}
\end{figure}

\textbf{Mixing} method (see Figure \ref{fig:mixing}) concatenates neutral speech recordings with other emotional speech recordings to create training samples that contain emotional intervals. Subsequently, we can apply supervision to each interval.

\end{document}